\documentclass[acmsmall]{acmart}
\usepackage{subcaption}

\AtBeginDocument{%
  \providecommand\BibTeX{{%
    \normalfont B\kern-0.5em{\scshape i\kern-0.25em b}\kern-0.8em\TeX}}}

\setcopyright{acmlicensed}
\acmJournal{PACMHCI}
\acmYear{2021} \acmVolume{5} \acmNumber{CSCW1} \acmArticle{16} \acmMonth{4} \acmPrice{15.00}\acmDOI{10.1145/3449090}

\received{June 2020}
\received[revised]{October 2020}
\received[accepted]{December 2020} 

\begin{document}

\title{Modular Politics: Toward a Governance Layer for Online Communities}

\author{Nathan Schneider}
\affiliation{\institution{University of Colorado Boulder}\country{USA}}
\email{nathan.schneider@colorado.edu}
\author{Primavera De Filippi}
\affiliation{\institution{Harvard University}\country{USA}}
\affiliation{\institution{CNRS}\country{France}}
\email{pdefilippi@cyber.harvard.edu}
\author{Seth Frey}
\affiliation{\institution{University of California, Davis}\country{USA}}
\email{sethfrey@ucdavis.edu}
\author{Joshua Z. Tan}
\affiliation{\institution{University of Oxford}\country{UK}}
\email{joshua.tan@magd.ox.ac.uk}
\author{Amy X. Zhang}
\affiliation{\institution{University of Washington}\country{USA}}
\email{axz@cs.uw.edu}

\renewcommand{\shortauthors}{Nathan Schneider et al.}

\begin{abstract}
Governance in online communities is an increasingly high-stakes challenge, and yet many basic features of offline governance legacies—juries, political parties, term limits, and formal debates, to name a few—are not in the feature-sets of the software most community platforms use. Drawing on the paradigm of Institutional Analysis and Development, this paper proposes a strategy for addressing this lapse by specifying basic features of a generalizable paradigm for online governance called Modular Politics. Whereas classical governance typologies tend to present a choice among wholesale ideologies, such as democracy or oligarchy, Modular Politics would enable platform operators and their users to build bottom-up governance processes from computational components that are modular and composable, highly versatile in their expressiveness, portable from one context to another, and interoperable across platforms. This kind of approach could implement pre-digital governance systems as well as accelerate innovation in uniquely digital techniques. As diverse communities share and connect their components and data, governance could occur through a ubiquitous network layer. To that end, this paper proposes the development of an open standard for networked governance.

\end{abstract}

\ccsdesc[500]{Human-centered computing~Collaborative and social computing theory, concepts and paradigms}
\ccsdesc[500]{Human-centered computing~Collaborative interaction}

\keywords{Governance; Institutional analysis and development; Online communities; Peer production; Platforms; Interoperability;  Standards}

\maketitle

\section{Introduction}

Many of the pressing social questions surrounding Internet technology and culture are questions of governance. \textit{Who should set policies that govern others’ behavior? What content should and shouldn't be allowed? What should be done with users' personal data? How should economic value be distributed?} Yet the governance tools available for online networks—such as social media groups, multiplayer games, and peer-production projects—are remarkably impoverished, typically relying on assigning unchecked power to admins and moderators~\cite{schneider-admins2021}. Such mechanisms as elections, boards, term limits, and transparent decision-making are norms for any incorporated entity in robust legal regimes, but to the extent that they do occur in online networks, they are costly to build and maintain, and they typically must be implemented by means extraneous to the feature-set of the network’s platform itself. Perhaps it is thus not surprising that monarchic or oligarchic practices emerge in online networks more easily and frequently than liberal or democratic ones~\cite{hill-wikipedia2013,frey-emergence2019,schneider-admins2021}.

We propose a strategy for addressing this lapse by modeling and specifying a governance layer for online networks with the following eventual design goals:

\begin{enumerate}
    \item \textbf{Modularity}: Platform operators and community members should have the ability to construct systems by creating, importing, and arranging composable parts together as a coherent whole.
    \item \textbf{Expressiveness}: The governance layer should be able to implement as wide a range of processes as possible.
    \item \textbf{Portability}: Governance tools developed for one platform should be portable to another platform for reuse and adaptation.
    \item \textbf{Interoperability}: Governance systems operating on different platforms and protocols should have the ability to interact with each other, sharing data and influencing each other's processes.
\end{enumerate}

To achieve these, we suggest the development of an open standard, along with supporting software libraries. We call our model ``Modular Politics.''

The model is preliminary and provisional, a starting point for future work. Modular Politics does not constitute a solution to aforementioned challenges such as content moderation or value distribution. No computational system can capture the embodied and cultural fullness of human governance practices; at best, computational tools can facilitate those non-computational processes. Nor do we argue for the superiority of some governance mechanisms over others. Rather, our proposal points in a direction of how users and platform designers might adjudicate challenges in creative and diverse ways. By proposing a pliable set of tools, we recognize that Modular Politics can allow the implementation of oppressive regimes as well as more democratic ones—but we suspect that pliability will result in greater accountability overall than what emerges from the autocratic administrative tools that prevail on social platforms today. Evaluating this suspicion would become far more possible with a laboratory like Modular Politics at hand.

Despite the near-ubiquity of monarchy and oligarchy on dominant social platforms, diverse tools for governance are as old as the social Internet ~\cite{ludlow-crypto2001}. These range from the experimental 1982 text-based game Nomic~\cite{suber-appendix1990} and the VOTEMGR software available for the early FidoNet bulletin board system~\cite{castillo-votemgr1991} to current cloud-based platforms like Election Runner that assist, for example, in elections for corporate boards and ``liquid democracy,'' which several Pirate Parties around the world use to build their political platforms~\cite{DBLP:journals/corr/KlingKHSS15}. 
Platforms such as Loomio~\cite{jackson-open2016} and Decidim support deliberation and community building as well as various voting mechanisms. The self-hostable multiplayer game Minecraft allows administrators to install ``mod'' plugins that can include tools for governance~\cite{frey-emergence2019}. A new wave of governance systems is emerging among distributed-ledger technologies, particularly surrounding the concept of decentralized autonomous organizations, or DAOs~\cite{alston-constitutions2019,reijers-now2018,defilippi-blockchain2018}. Yet most users of social platforms do not have the means of experiencing the diverse range of possible governance technologies. Modular Politics seeks to make a wider range of governance possibilities available to users and researchers—possibilities that have yet to be meaningfully tested or even tried.

What follows is a conceptual overview intended as a prologue for future theoretical, technical, and empirical work, not a specification or theory. For example, this overview does not consider matters such as security and database structures that will be vital at the design stage, nor does it present the kinds of empirical claims that we expect to develop and test with an eventual prototype.

First we will review other approaches to designing online governance regimes that motivate our design goals. We will then present a preliminary sketch of the basic features that we expect Modular Politics to include, followed by some examples of how our model might work in practice. Next, we consider strategies for implementation, proposing the development of an open standard, and conclude with a discussion of future research directions enabled by Modular Politics.

\section{Background}

The governance of online communities has long been a topic of study within the fields of computer-supported cooperative work, social computing, and computer-mediated communication.
Early studies examined discussion groups hosted on newsgroups~\cite{kollock-managing1996}, mailing lists~\cite{hyman2003twenty,cubbison1999configuring}, Internet Relay Chat clients~\cite{reid1991electropolis}, or bulletin board systems~\cite{stone1991will}.
In the past several decades, researchers have gone on to study the governance of large-scale peer production communities such as Wikipedia~\cite{forte2009decentralization} and open source software projects~\cite{schweik2012internet}, online multiplayer games such as League of Legends~\cite{kou2013regulating}, as well as large social media platforms such as Facebook~\cite{klonick2017new}.

\subsection{Problems in Online Communities}
Participants in early discussion and gaming communities complained about reoccurring issues of unwanted behavior, including spam and harassment~\cite{dibbell-rape1993}, as well as conflicts with other members, sometimes called ``flame wars''~\cite{talja2004field}.
Today, online harassment and trolling have gone mainstream with the ubiquity of social interaction online, resulting in evidence that nearly half of Internet users in the U.S.~\cite{pew,datasociety} and nearly three quarters of people who play online games have experienced some form of online abuse~\cite{league2019free}. 

During the 1990s, software peer production communities~\cite{benkler-wealth2006} began to form, with famous examples such as the GNU/Linux operating system~\cite{schweik2012internet, benkler-coase2002} and the collaborative encyclopedia Wikipedia~\cite{forte2009decentralization}.
Like discussion communities, peer production communities must contend with unwanted behavior such as vandalism~\cite{geiger2010work} and sockpuppetry~\cite{solorio2013case}. 
However, peer production communities often maintain stricter standards for contribution.
This can result in barriers for newcomers, including having their first contribution met with no response, impolite responses, or responses that are too complex for them to understand~\cite{steinmacher2015social}.
Peer production communities also may have certain productivity goals, and repetitive conflicts such as ``edit wars'' on Wikipedia~\cite{kittur2007he} or other intransigent disputes~\cite{im2018deliberation} can reduce productivity.

Communities face additional governance challenges as they begin to grow in size~\cite{ducheneaut2007life,halfakerrise2013}.
In the case of discussion groups, which tend to have only one or a small number of moderators, those moderators complain of time-consuming labor needed to filter out content~\cite{cubbison1999configuring}, as well as burnout from the emotional toll of constantly encountering toxicity~\cite{dosono2019moderation}.
To manage an increasing volume of content, some platforms have invested in machine learning tools to automatically remove unwanted content~\cite{perspective}. While spam today is mostly dealt with automatically, mitigating other forms of unwelcome content such as misinformation and harassment is harder to automate because of their contextual nature~\cite{binns2017like,hosseini2017deceiving}. 
With a growing number of community members, many may be unaware of established norms~\cite{kiene2016surviving}. While some communities resolve these problems with more rules governing participation~\cite{butler2008don}, they also run the risk of turning away newer members~\cite{kraut2012challenges,steinmacher2015social,halfakerrise2013}, resulting in low conversion rates of casual users to core contributors~\cite{panciera2009wikipedians} and entrenched leadership~\cite{shaw2014laboratories}.

The problem of scale reaches new magnitudes when it comes to massive, centralized social media platforms, such as Facebook or Twitter, where much of online social activity resides today.
Governance decisions by major platforms, such as the early decision by Facebook to enforce a real-name policy~\cite{haimson2016constructing}, have tremendous ramifications for society given the size of their user base and the inability for many people to opt out of such platforms to participate in society. 
As they have grown, platforms have become the ``new governors'' of an increasingly digital public sphere~\cite{klonick2017new}, and the governance that platforms provide have become a major component of their product offering~\cite{GillespieTarleton2018Coti}. 
In that regard, platforms have been criticized for their lack of transparency around governing~\cite{santaclara}, the slow response to emerging user problems such as harassment~\cite{geiger2016bot}, the outsourcing of content moderation to large teams of contractors facing poor working conditions~\cite{roberts2019behind}, and the lack of user input into governance decisions beyond low-level flagging~\cite{crawford2016flag}. Along multiple dimensions, the governance mechanisms currently available do not appear up to the task of meeting the pressing problems that social platforms face.


\subsection{Governance Approaches in Online Communities}

Recognizing the importance and ubiquity of these problems, researchers have taken interest in how communities tackle them. The findings from such work have informed our approach in what follows.

The largest social media platforms have attracted significant attention for their responses to governance problems, with a particular focus on the broader societal impacts of their governance choices~\cite{GillespieTarleton2018Coti,wu-attention2017,bonneau-democracy2009,bucher-algorithmic2017,roberts2019behind,gray2019ghost,phillips2015we}.  
The literature tends to take a critical posture, highlighting the marginalization of user input~\cite{lampe2004slash,crawford2016flag,matias2016civic} and focusing on how their profit motive results in governance decisions that are not in the interests of users~\cite{egelman2013my,nouwens2020dark}.
Analyses of centrally governed communities become more sympathetic as those communities become smaller and flatter; for instance, the group of researchers that produced \textit{Building Successful Online Communities}~\cite{kraut-building2011} provide a well-organized overview of insights into the problems these communities face and strategies for solving them.

Closer to the democratic ideals of early Internet evangelists~\cite{yar2014virtual}, self-governing communities occupy several seminal studies of online culture, as in research on the LambdaMOO and MicroMUSE platforms~\cite{dibbell-rape1993,10.1111/j.1083-6101.1996.tb00185.x,smith2002problems}, in which crisis events galvanized virtual communities in ways that spurred action and even led to formal governance schemes. 
Several important online organizations employ remarkably democratic governance models. These include the World Wide Web Consortium, the Wikimedia Foundation, the Apache Software Foundation, the Debian operating system, and, after a recent restructuring, the developers of the Python programming language. 
Among these large-scale, more-or-less participatory organizations, the projects of the Wikimedia Foundation have attracted the most governance research, English Wikipedia in particular~\cite{forte2009decentralization,keeganCodeWrittenSand2015,butler2008don,heaberlin2016evolution}. 
Other large platforms have dabbled with more circumscribed experiments in democracy~\cite{Ullyot2009,kou2013regulating}, or seen such experiments emerge from the user community~\cite{jhaver2019human,matias2015reporting,geiger2010work, fandigital}.  While these more democratic examples play crucial roles in Internet infrastructure, their governance is more an exception than the rule. They also tend to rely on bespoke processes and tools, rather than easily replicable ones.

Alongside such procedural democracies, researchers have investigated the many platforms that rely on more decentralized structures.
Examples of this include USENET~\cite{kollock-managing1996}, mailing lists, Facebook groups, Reddit communities, file-sharing networks~\cite{loban-rhizomes2004,harrisInstitutionalSolutionsFreeriding2018}, the Tor network~\cite{hardyReputationInternetBlack2016}, blockchain protocols~\cite{reijers-now2018,pazaitis-blockchain2017,defilippi-blockchain2018}, and, of course, the World Wide Web itself \cite{HessCharlotte_1995}.  
Independent units of these platforms may be seen to adopt many types of governance styles, but considered as populations of populations, they can be characterized by the freedom users have to exit any one community for any other---that is, they favor ``exit'' over ``voice''~\cite{hirschman-exit1970}.     
Member communities of these platforms are most likely not represented in platform-scale decisions, but through grassroots collective action, campaigns have exercised the platform-scale influence they formally lack ~\cite{matias-going2016,centivany2016popcorn}. 

This self-organization is especially evident in their ability to develop free and flexible ``off-platform'' software tools for addressing the governance challenges they share in common, whether via general tool libraries~\cite{jhaver-humanmachine2019,chandrasekharan2019crossmod}, templating mechanics~\cite{fiesler2018reddit,butler2008don}, or even community-led randomized experiments~\cite{matias2018civilservant}. 
For example, in lieu of centralized platform action on unwanted behavior, community members and end users have piloted new governance designs to dissuade norm violators~\cite{matias2019preventing} and protect against harassers~\cite{mahar2018squadbox, geiger2016bot}, as well as to facilitate reconciliation between conflicted members~\cite{schoenebeck2020drawing} and provide community support for victims~\cite{blackwell2017classification}.
Self-organization is also evident in the ability of communities to develop and sustain widely held informal norm systems~\cite{castronovaResearchValueLarge2006a,cavenderInternalReinforcementCooperative2013,rossSocialScientificFramework2014a,strimlingEmergentCulturalDifferences2018}.
Research on online criminal communities is an especially rich source of insight into the potential of norm-heavy governance systems ~\cite{sadiaHonorThievesCommon2013,hardyReputationInternetBlack2016}. 
Complementing the research community's overall faith that technology \textit{can} play a supporting role in online governance, researchers agree that technology alone does not define a governance system or determine its success. 
Cases abound demonstrating the importance of a healthy culture to a platform's continued existence, regardless of its governance technologies~\cite{cosleyHowOversightImproves2005,kiene2016surviving,gillespie-politics2010,grimmelmannVirtuesModeration2015,seering-moderator2019,caiWhatAreEffective2019,kieneManagingOrganizationalCulture2018}. 

Prior research suggests that although sophisticated and democratic governance is possible in online communities, it is difficult and rare. The tools available to communities far more often leave users to develop governance practices in an informal, \textit{ad hoc} fashion.  While we cannot claim that improved governance tooling alone would solve the vexing problems of social platforms, there is reason to believe that an improved framework for experimentation with governance models could generate pathways for improvement.

\subsection{Institutional Analysis and Development}

Questions of governance design are of great interest outside of the sciences of information and technology, particularly in the social sciences. 
Yet social scientists have generally been reluctant to move beyond classification and characterization into design. 
Plato and Aristotle each offered taxonomies of political systems, employing such labels as monarchy, oligarchy, and democracy; their emphasis on categorizing macro-level structures---assuming a closed set of possible governance forms---continues to hold sway, even as more recent typologies of comparative politics introduce multi-dimensional complexity so as to reflect the interplay of diverse institutions and practices found in modern governments~\cite{blondel-comparative2014,lindberg2014v,elkins2014constitute}. 
Quantitative comparative political scientists have a much finer-grained view of governance design, but still restrict their focus largely to national-scale governance, exploring questions focused on political organization~\cite{Scarrow_Webb_Poguntke_2017,Janda_1980}, political processes~\cite{lijphart_patterns_1999, gallagher_comparative_2005, voigt_positive_2011}, and political economy~\cite{Landman_2000}. Governance design approaches in other social sciences are open to smaller-scale subjects, but have tended to adopt the convenience of assuming a powerful designer, such as a founder, CEO, or administrator. 
This describes the dynamical systems~\cite{forrester1997industrial} and cybernetics~\cite{beer1972brain,steenson_ch5_2017} branches of complex systems theory, which were both introduced as social design paradigms. Design approaches in the management literature on organizational processes recognize managers as a type of designer~\cite{devreedeUsingGSSDesign2000,beck2001manifesto,maloneToolsInventingOrganizations1999}. 
Economics has a rich literature on market-mechanism design, which has been successful enough to influence contexts from voting systems to spectrum allocation~\cite{borgers-introduction2015,phelps-evolutionary2010,jackson2014mechanism,lalley2018quadratic}. 
Alternatively, participatory design frameworks are premised on the importance of broadening the design process to include all types of stakeholders~\cite{kensing1998participatory,asaro2000transforming}. 
This participatory tradition has been formative to the development of CSCW and modern HCI~\cite{bannon2011lest}. 

Even allowing this range of examples, we conclude that social science has given surprisingly little attention to general and systematic techniques that a lay practitioner can use to craft governance institutions. 
Fortunately, growing interest in natural-resource management and skepticism toward rigid market/state dichotomies have led to rapid development in a form of institutional inquiry oriented around the design of governance across a wide range of social contexts, one beginning with participant agency rather than macro-level structure. 
This effort has been centered at the Ostrom Workshop at Indiana University, founded by political scientists Elinor and Vincent Ostrom. Their Institutional Analysis and Development (IAD) framework, though focused on analysis, is well suited to the challenge of design. 
In fact, it was aspirationally intended to become a design tool: in IAD's first instantiation, the ``D'' stood for ``Design''~\cite{ostrom1998institutional}.

An important feature of IAD is its sensitivity to the diversity and complexity that tend to emerge among actually existing human governance regimes~\cite{ostrom-understanding2006,ostrom-designing1995,mcginnis-networks2011}. Many of the cases the Ostroms and their colleagues studied were from among longstanding, Indigenous, and non-Western societies that operated quite differently than the usual subjects of economics and political science research.
This commitment to empiricism across diverse settings has put IAD at odds with the theory-driven approaches that have dominated the sciences of governance, a tension captured by ``Ostrom's Law,'' which asserts that ``a resource arrangement that works in practice can work in theory''~\cite{fennell2011ostrom}. 
IAD's empirical orientation, and its consequent openness to not-yet-theorized governance forms, has equipped it to capture institutional complexity. Empiricism also makes IAD suitable for a more hands-on approach to institution design, a divergence from the tendency to privilege theory in the governance analyses of economics. This commitment to the experiential component of design was important to the Ostroms, who emphasized the craft of institution design and likened its practitioners to artisans~\cite{ostrom1980artisanship}.  

There are, of course, shortcomings to applying IAD to online governance. Although some have argued that IAD is basically aligned with contemporary critical discourse around power and culture~\cite{Shrestha_Ojha_2017}, a credible case can be made that IAD's commitments to rational choice and methodological individualism---consequences of its alignment with quantitative behavioral science---make it unprepared to adopt ``a  broader  concept  of  human agency, contextualise and emphasise the importance of history, and look at the social construction of rationality, interests and identity''~\cite{Mollinga_2001}. IAD has also not yet been adapted into a computational governance environment at the scale we propose. Nevertheless, we find IAD to be a promising basis for modeling agency in a computational context.

IAD's basic unit of institutional structure, the ``action situation,'' is a game-like confluence of conditions in which participants make individual choices and collective decisions~\cite{mcginnis-introduction2011,ostrom-governing1990}. IAD normally allows for participant involvement in shaping the core rule-set of an action situation in their shared institutions~\cite{frey-this2019}. 
To represent more complex institutions, IAD represents action situations as becoming nested and interlinked into systems of larger action situations that, when combined with a set of users and other context become an ``action arena''~\cite{ostrom-understanding2006}. 
The Ostrom Workshop community went on to develop a general concept of polycentricity for effective action arenas, which describes the even more complex interplay of multiple institutions that the Ostroms observed in the field~\cite{ostrom1999polycentricity,ostrom2010beyond}. 
The polycentric frame can encompass nested, overlapping, and competing divisions of power that interact to shape participants' action situations. 
For example, how storm water runs in and out of a city may depend on activity from many levels of government (federal, state, county, and municipal), many divisions of government (planning departments, parks departments, and waste management departments), many modalities of government (legislatures, courts, and law enforcement), and many stakeholders in government (officials, NGOs, and citizens).
A recurrent finding of IAD research is that governance organizes the complexity of each level to match the complexity of the social-ecological system with which it interfaces~\cite{ostrom-designing1995}. 

Previous research about online governance has adopted the IAD framework~\cite{HessCharlotte_1995,kollock-managing1996,viegasHiddenOrderWikipedia2007,forte2009decentralization,kraut-building2011,schweik2012internet,silberman-reading2016,frey-emergence2019,reijers-now2018,frey-this2019,seidel-democratic2019}. We argue for a further embrace of the IAD approach in online governance design, adopting IAD's emphasis on the agency and creativity of participants. 
The notion of governance as interlinked action situations is ideal for designing complex governance environments from a user-centric point of view, rather than presuming an idealized and highly constrained agent (as in game theory) or abstracting away participant agency by focusing on structure (as in much comparative political analysis). 
Still, IAD scholars have yet to rigorously formalize the precise nature of these aggregations and their linkages: what specific ties are permitted between games, or how they come to faithfully represent a real-world institution~\cite{mcginnis-networks2011}. 
The approach we propose could be described as inviting communities to formalize the action situations that constitute their own action arenas, which communities can thread together across polycentric networks.

\subsection{A New Approach}

Given the bewildering variety of governance institutions and their complex relationship with culture, an online governance paradigm should enable communities to built nested assemblies of action situations. Doing so requires an approach to governance systems that is \textit{modular}, so that users can compose systems of nested systems from the bottom up, following IAD's nested conception of institutions. Elinor Ostrom conceptualizes this nesting in terms of Arthur Koestler's idea of the ``holon,'' which is when ``a \textit{whole} system is part of a system at another level''~\cite{ostrom-understanding2006}. The standardization that modularity requires should be in balance with \textit{expressiveness} sufficient to implement highly diverse, culturally heterogeneous systems that IAD's global perspective highlights. Finding this balance may involve compromises that limit expressiveness somewhat to ensure modularity and other design goals. 

As communities struggle with changes in their user base or operating environment, they benefit from being able to quickly iterate through targeted institutional changes. 
This process will be easier if communities are able to borrow and adapt working solutions from each other, and engage with the facility in the ``institutional artisanship'' that the Ostroms imagine~\cite{ostrom1980artisanship}. 
To facilitate this, a modular governance paradigm should also ensure that its components are \textit{portable} across contexts. Successful modularity will facilitate portability by requiring the components of a system to have certain standard features.

Online communities are networked communities. Participants may share hyperlinks with each other from across the Web, or leverage a following on one platform (e.g., on Instagram) to achieve their goals on another platform (e.g., crowdfunding on Kickstarter). If a governance system is to encompass the polycentricity of life online, it must be \textit{interoperable} across platforms, able to interact with far-flung data sources and interactions. Modularity, again, can help facilitate this by ensuring that the components of disparate implementations share common interfaces.

With Modular Politics, we seek to outline a paradigm that empowers online communities to assemble discrete, programmable building-blocks into original and complex governance regimes that reflect their aspirations and needs. The model we describe is capable of implementing governance mechanisms, connecting them into higher level constructions, organizing information, and managing resources. Tools created for one community can be adopted in others. The model also supports interaction among distinct communities and across networked platforms.

\section{The Model}

In this section, we present a preliminary overview of how the Modular Politics model might operate, drawing on the four design goals outlined above as well as the IAD research paradigm. We employ a set of specialized terms with meanings specific to this model, whose initial letters we capitalize to distinguish our usage from colloquial meanings. These terms serve to establish a technical vocabulary for the model, but in specific implementations, developers may replace them with a user-facing vocabulary more appropriate to their world-building context.

We present the model alongside a hypothetical use-case to help illustrate how each component might work. Consider a fictional multiplayer virtual game, Guilded Age, in which teams known as guilds compete with each other to acquire resources and earn points by building monuments. The game operates on a federated network structure in which each guild resides on a separate user-administered server across a shared protocol, similar to games such as Minecraft or World of Warcraft. Guilds may be as large as 100 players each, and they make collective decisions about strategies for resource acquisition, architecture, and construction. The rules of the game are fixed, but the guild associations can choose governance structures for themselves from a large pool of available components. Players are free to choose which guild they want to join for a given round—or they can form their own and attempt to attract others to join. At the start of each round, server administrators assemble a set of governance structures for the guilds they want to form, and other players can use a guild-comparison tool to examine the governance regimes and member attributes in each guild before choosing which guild to join. These guilds can take a variety of forms. Guild founders with high levels of reputation in the game end to succeed in attracting players to join autocratic guild structures, as past accomplishments imbue trust in those founders’ leadership. A diverse range of options for governance thus emerges, as does an ever-growing variety of creative governance techniques.

\begin{figure}
  \centering
  \includegraphics[width=\columnwidth]{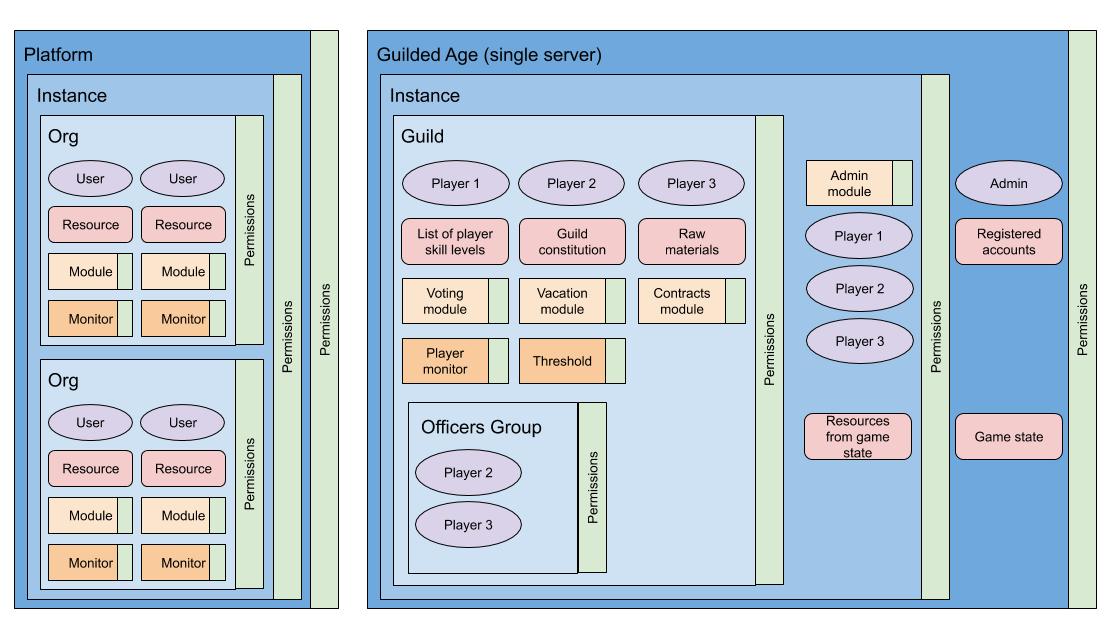}
  \caption{A schematic implementation of the system described throughout this section. Here, a single server running Guilded Age also runs an Instance of Modular Politics. The Instance translates game data into assets recognizable by different Orgs and Modules. Finally, guilds in Guilded Age correspond to Orgs, and each guild may install a number of Modules and track a number of Resources. An Org may have a sub-Org, e.g. a guild may have a sub-group of officers with special permissions.}~\label{fig:guild}
\end{figure}

\subsection{Platforms, Instances, and Orgs}

Modular Politics is not a standalone system but must be implemented within an underlying \textit{Platform}---perhaps a game, social network, or blockchain protocol. An implementation of Modular Politics is called an \textit{Instance}. The Instance defines the interface between Modular Politics and the underlying Platform---specifying the entities from the Platform that have a role in the governance process and the actions that they can take on the Platform (Figure \ref{fig:guild}). In defining the Instance, Platform operators determine who has access to Modular Politics and what it has the power to govern. The Platform in Guilded Age is the game's software, which runs on user-administrated servers across the network. An Instance of Modular Politics is embedded in the software and thus runs on each server concurrently. The game software defines certain ground-rules for each Instance and associates events in Modular Politics with events in gameplay.

Within an Instance, governance occurs through computational domains called \textit{Orgs}. Orgs provide the institutional context for what IAD calls action arenas in Modular Politics. They are semi-autonomous governance environments, of which an Instance can have many. These are the tools by which Guilded Age players begin to craft the governance of their guilds. Org creation allows for mimicking a variety of organizational types, such as companies, clubs, factions, boards, and more. An Instance’s Org or Orgs may be specified by Platform operators in advance; the system’s participants may also be allowed to create and join their own Orgs voluntarily. In Guilded Age, server administrators can choose either option for their users. Orgs can be nested or formed in parallel; when one Org is created within another, its members can include any members of the parent Org. Orgs are constrained only by the fact that their members are also subject to any other Orgs within which they are members. A Guilded Age guild can thus have sub-groups within it. This accords with the practice of ``nesting,'' which Elinor Ostrom observed as a recurrent pattern in commons-based governance.

\subsection{Modules}   
Following the IAD concept of modular action situations, participants in Modular Politics take action through computational \textit{Modules}. Modules are configurable software packages that can alter the behavior of processes that run on an Instance or on particular Orgs within it. A Module's editable configuration options are called \textit{Policies}. Compatible Modules can be combined to form more complex Modules. Modules, as their name suggests, are critical for achieving the design goals of modularity in this model, as well as arbitrary expressiveness; rather than prescribing specific governance systems, Modular Politics enables a bottom-up approach to system creation.

Guilded Age administrators, for instance, might choose to define a certain set of Modules for their players, or they might load a Module to the Instance that provides players with a process to add Modules and Orgs or alter Policies during gameplay. Administrators and players can adopt Modules—or packages of Modules—from third-party repositories or create their own.

The designs for Modules may include typed inputs and outputs, which help facilitate conjoining Modules into ever more complex ones. The inputs and outputs can be Users, Resources, Orgs, or other types of data; inputs and outputs can also come from other Instances (or non-Modular Politics systems) through API calls, if Policies allow. A Module that outputs a decision can provide input for a Module on a separate Instance that translates the decision into a Policy change. In Guilded Age, the guild on one instance might use this method to form a contract with another guild on another server; if one guild violates the contract, the Module automatically pays restitution to the other server's guild. This kind of interoperability is essential for supporting what IAD regards as the inevitable emergence of polycentricity.

Modules are also transparent. Their underlying source code are accessible to anyone who can view and interact with them, so users can audit the mechanics of their behavior. This includes Modules on separate Instances being accessed through API calls; the API enables remote access to Module source, both for inspection and adoption, although a Module's Policies may have restricted permissions. Because they are all on a common network through APIs, Guilded Age guilds can see which Modules are running on opposing guilds. This buttresses the model's design goals of interoperability and portability.

Modules can have more than one functionality depending on how they are called. For example, a single Module for petitions can be called to either create a new petition, sign an existing one, or query a petition's status. Policy configurations specify these various behaviors. A Module for an elected guild leader will include a Policy specifying the length of the leader's term and another specifying the threshold of support required for the leader to be elected. To operate like this, Modules must be able to interact with, and change the states of, representations of participants and Platform data.

\subsection{Entities}

An Instance can include two types of entities: 

\begin{itemize}
    \item \textit{Users} are causative agents, such as humans or bots
    \item \textit{Resources} are objects or actions that Users can affect
\end{itemize}

Elinor Ostrom stressed the importance of clarifying the bounds of the membership and the domain of a self-governing community. In Modular Politics, each entity has a state in the Instance and contributes to the state of the Instance as a whole. These entities can also become members of the Orgs in their Instance, which makes their states subject to the Modules that govern any Org they join. Entities may be assigned to particular Orgs by the Platform operator, or through voluntary agreements among Users. Users can assign that Resource to membership in that Org. If the Org is nested within another Org, entities must belong to or be added to the parent Org in order to join.

Each player on a Guilded Age server is a User; the server administrator decides the criteria for joining. The Guilded Age software correlates Resources in the Modular Politics Instance with the elements of gameplay, such as tools, skills, and raw materials for monument-building. The administrator may choose to include a Module that allows players to create and layer their own Orgs.

Thus, the Instance configuration determines the relationship between Users and Resources in Modular Politics and the Platform. This relationship includes an identity management system for Users, as well as establishing any correlation between Resources and their manifestation on the Platform. To achieve the design goals of expressiveness and portability, it should be possible to connect virtually any kind of computational object a Platform might have into a User or Resource in Modular Politics.

\subsection{Monitors}

\begin{figure}
  \centering
  \includegraphics[width=.8\columnwidth]{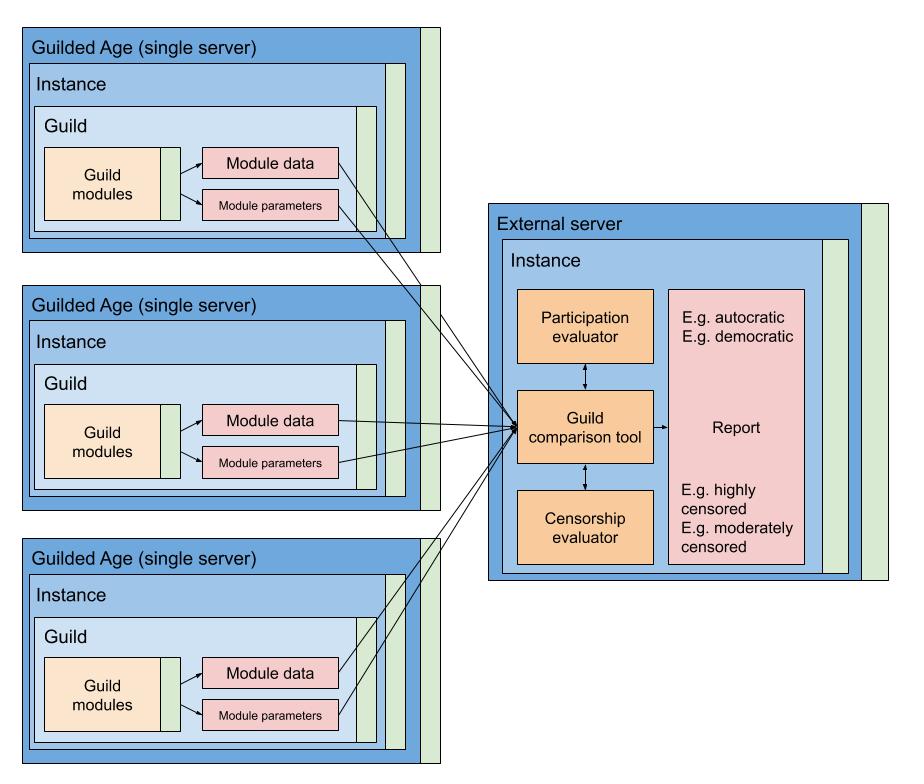}
  \caption{An example of a Monitor working across different Instances of Modular Politics. Continuing the Guilded Age example, here an external service queries a set of Modules shared between a set of guilds and then compares those guilds based on the data obtained from those Modules.}~\label{fig:model}
\end{figure}

\textit{Monitors} are ``read-only'' Modules that provide feedback by evaluating specific conditions according to consistent criteria. Their analyses can take place over a certain duration or based on a snapshot at the time the Monitor is queried.

Monitors can gather data from throughout the system, such as by querying the states of specific Modules, Users, and Resources, or by querying the Policies that specify how a Module behaves. In Guilded Age, a Monitor might be able to query the tools and skills that players in one's guild possess and compare them to players in other guilds (Figure~\ref{fig:model}). Another Monitor might observe the rate at which each guild is building monuments. A guild could also install a Module that, when its monument-building falls behind a certain percentile compared to other guilds, abolishes the breaks from working that players otherwise take.

Monitor outputs can consist of multiple data types, such as a binary value (whether or not certain criteria are satisfied), a fractional value (such as a percentage score based on partial satisfaction of criteria), or an array (multiple data points on a given query). These outputs provide information on the states and behaviors of entities, Instances, Orgs, or Modules. They can also acquire considerable complexity by integrating diverse data sources or aggregating multiple other Modules through nesting. As a special type of Module, Monitors are intrinsically composable and thus contribute to the larger objective of modularity. A sophisticated Monitor could make broad claims about the nature of an Org, akin to indexes that evaluate the democratic health of governments~\cite{giannone-political2010,diamond-quality2004}. Because Monitors’ source code and Policy configurations are available for inspection, their analytic methods should be in principle transparent.

\subsection{Permissions}

Access and authority in Modular Politics may not be equally distributed. Permissions are part of the configuration Policies for any Instance, Org, and Module. These permissions specify what actions Users can take in a particular environment. A User can create a new Org within an Instance, for example, only if the Instance’s Policies grant that User the ability to do so. A Module can only query the state of a given Org if the User triggering that Module has obtained the necessary permissions to query that Org. Ultimately, all permissions in Modular Politics derive from those that the platform administrators specify at the level of the Instance. A Guilded Age server might grant all users the right to use an Org-creating Module, for instance, or it might reserve that right for players who have been elected as masters of their guild.

In addition to specifying permissions within a given Instance or Org, a Module’s Policies state whether it allows external API calls from other Instances, which may be on the same Platform or another. Such permissions are necessary in order to ensure that the goal of interoperability is handled responsibly. Moreover, a Module might be configured to allow for certain functions to be called only by specific Users, or it might allow access based on Users’ membership in certain Orgs or Instances. Since Guilded Age is intrinsically a networked game, the software requires API access to basic functions, though individual guilds can decide whether it allows activities like interactions with players in different guilds.

Restrictions defined in a particular Instance or Org are automatically applicable to every Org created within it; permissions are inherited hierarchically. For example, if an Instance forbids external API calls, none of its Orgs can accept external API calls from Instances on other Platforms. Conversely, if an Instance is configured to accept external API calls, it will be possible for external Modular Politics Instances to interact with entities and Modules in the Instance—assuming that such interaction isn’t elsewhere restricted.  One Platform’s Instance, consequently, could provide governance services to other Instances through APIs.

In Guilded Age, a guild might hold certain tools in an Org that only players with a certain rank in the guild can join. A Monitor, also, might be able to access certain information only if the player triggering it holds a given rank. And the Module that sets the players' rank may rely on the votes of only high-ranked players to carry out a promotion.

\subsection{Further Configuration}

\begin{figure}
\begin{subfigure}{.5\textwidth}
  \centering
  \includegraphics[width=.8\linewidth]{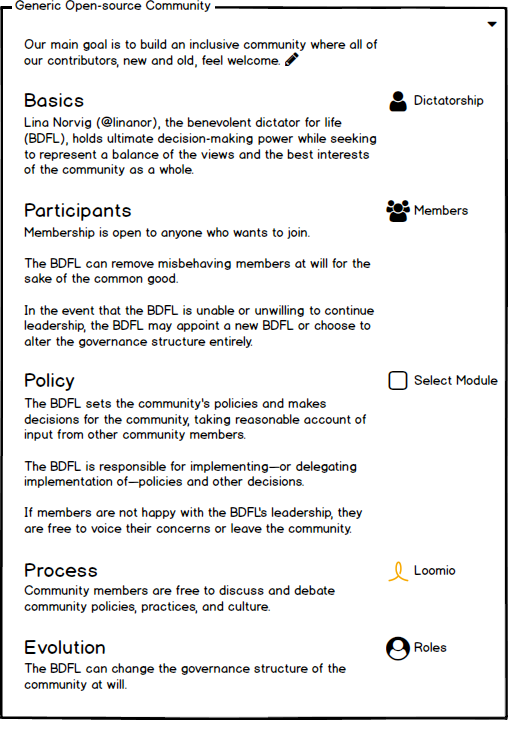}
  \caption{}
  \label{fig:config1}
\end{subfigure}%
\begin{subfigure}{.5\textwidth}
  \centering
  \includegraphics[width=.8\linewidth, trim=0 450 0 0, clip]{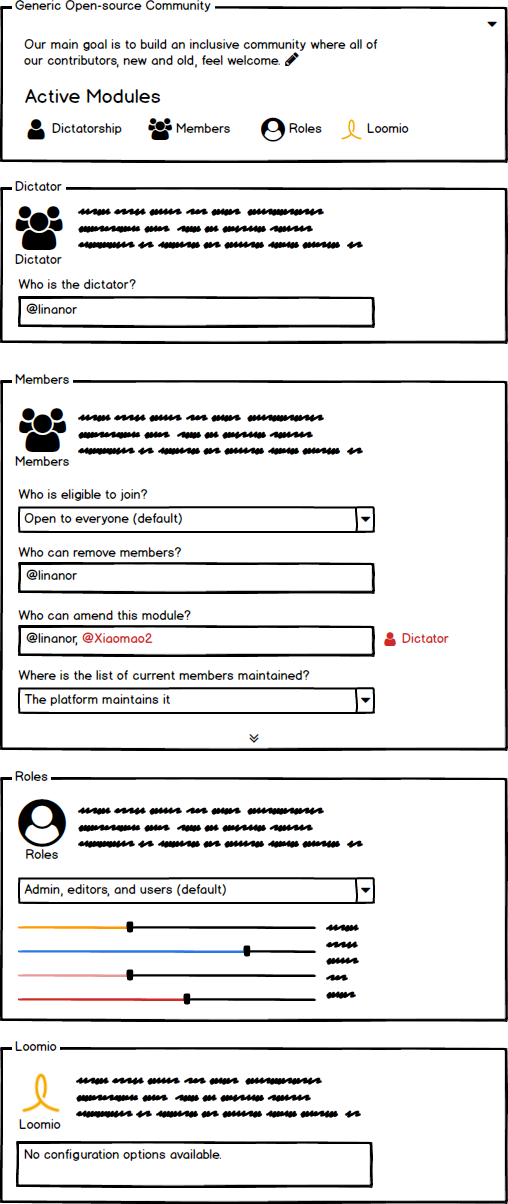}
  \caption{}
  \label{fig:config2}
\end{subfigure}
\begin{subfigure}{.5\textwidth}
  \centering
  \includegraphics[width=.8\linewidth, trim=0 250 0 0, clip]{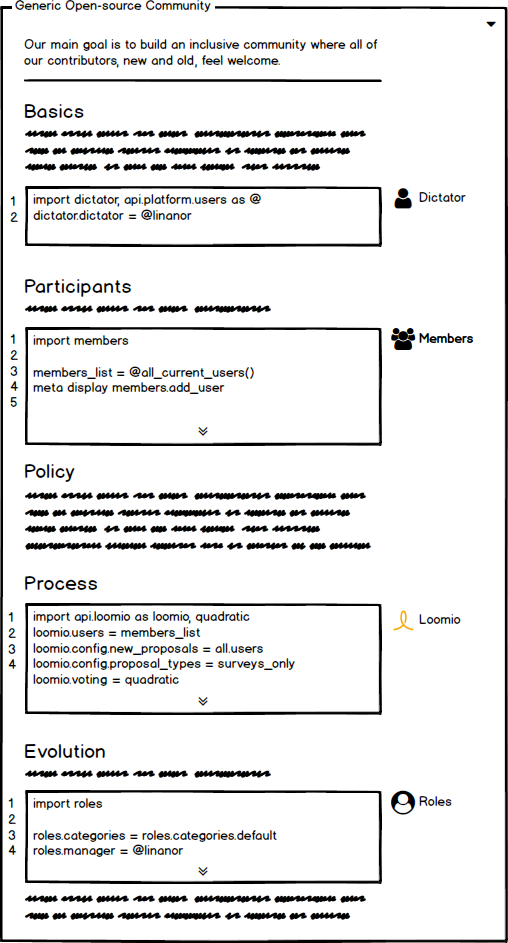}
  \caption{}
  \label{fig:config3}
\end{subfigure}%
\begin{subfigure}{.5\textwidth}
  \centering
  \includegraphics[width=.8\linewidth]{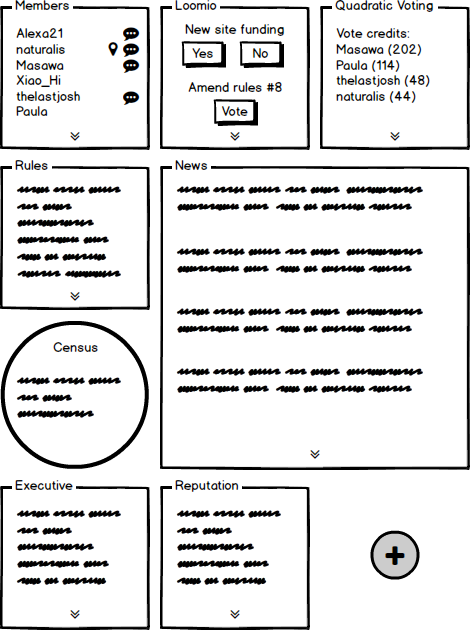}
  \caption{}
  \label{fig:config4}
\end{subfigure}
  \caption{A set of example configuration interfaces for Orgs and Modules. (a) depicts a text-only representation that approximates an Org's textual constitution or code of conduct, (b) depicts only the actual configuration options of Modules along pre-defined parameters, (c) interpolates text with code snippets in a notebook-style interface, and (d) presents a possible end-user interface produced by the underlying Modules. Note that configuration Policies could also be made through visual programming or a simplified palette of options.}~\label{fig:config}
\end{figure}

Just as IAD emphasizes the need for participants to shape their governance environments, Modular Politics aims to provide wide latitude for configuring a governance environment without requiring extensive technical knowledge. Platform operators and their Users, if they have the necessary permissions, can create and customize their own governance systems. Configuration of these systems occurs at the level of the Instance, Org, and Module. Some Instances may have all their Orgs and Modules predefined by the Platform operators, with no opportunity for change or configuration by Users; other Instances may enable Users to modify the arrangement of Orgs and Modules as they choose. Operators can specify a fixed set of Modules available to their Users, or they can allow users the flexibility to acquire their own Modules. Discrete Platforms and communities using Modular Politics should be able to share, copy, and modify each other’s Modules and Policies through public or private code repositories. Meeting the design goal of portability will be essential for accelerating governance innovation among users.

Configuration is essential in Guilded Age, as it is what enables guilds to define their own structures and rule-sets. Some guilds will allow configuration only by administrators, while others will see value in encouraging self-governance and creativity among players.

\begin{figure}
\begin{subfigure}{.5\textwidth}
  \centering
  \includegraphics[width=.8\linewidth]{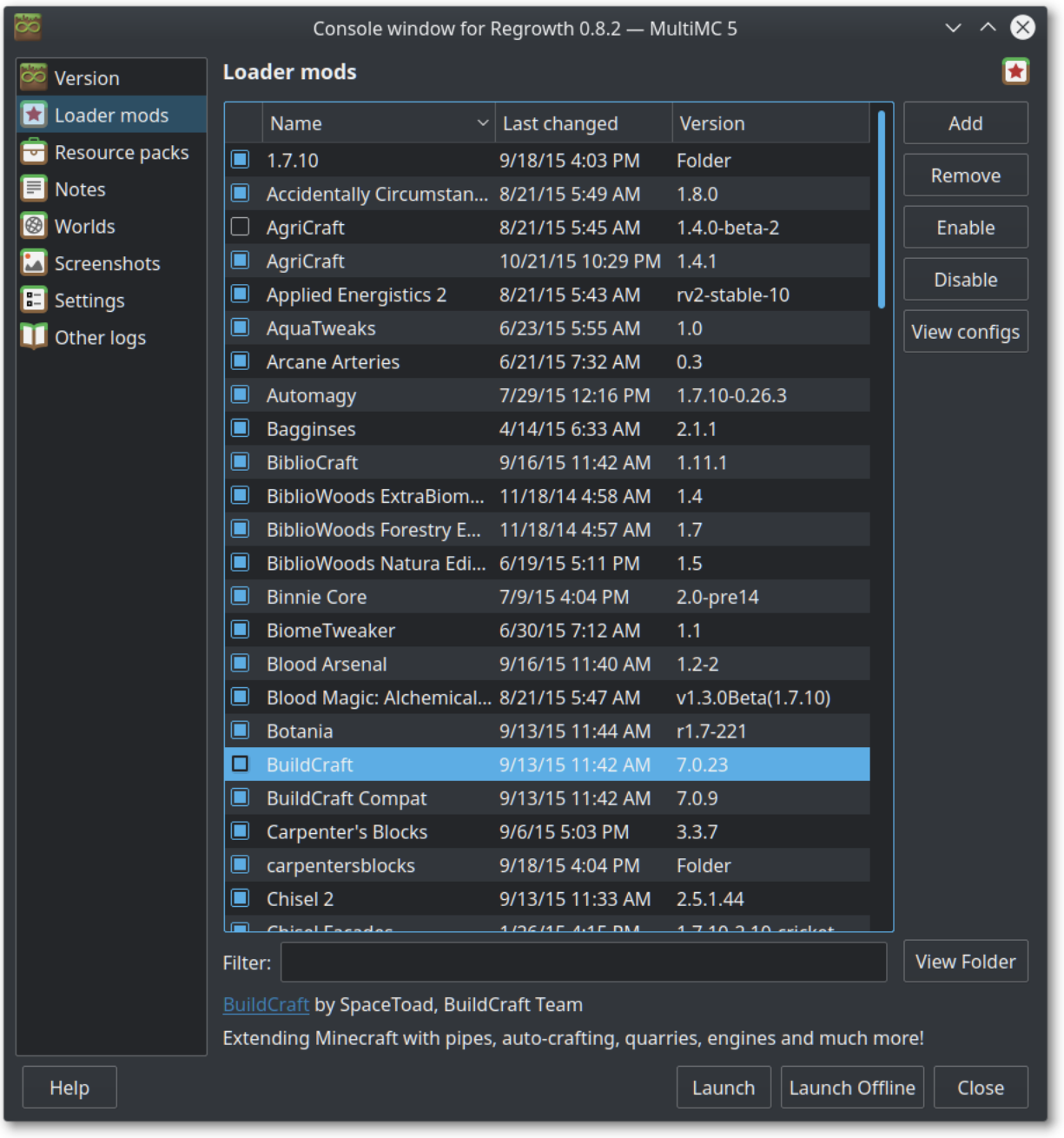}
  \caption{}
  \label{fig:multimc}
\end{subfigure}%
\begin{subfigure}{.5\textwidth}
  \centering
  \includegraphics[width=.8\linewidth]{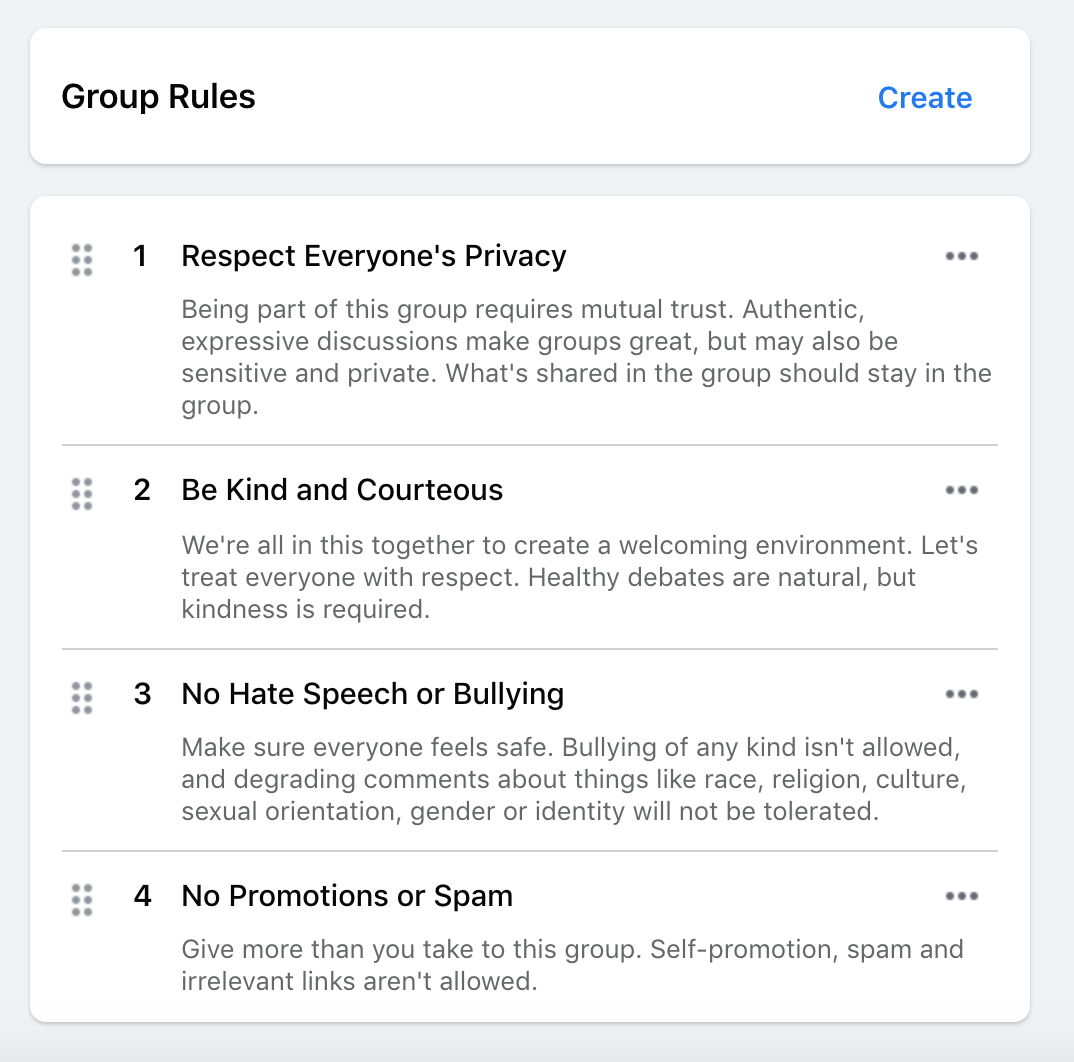}
  \caption{}
  \label{fig:facebook}
\end{subfigure}
\begin{subfigure}{.5\textwidth}
  \centering
  \includegraphics[width=.8\linewidth]{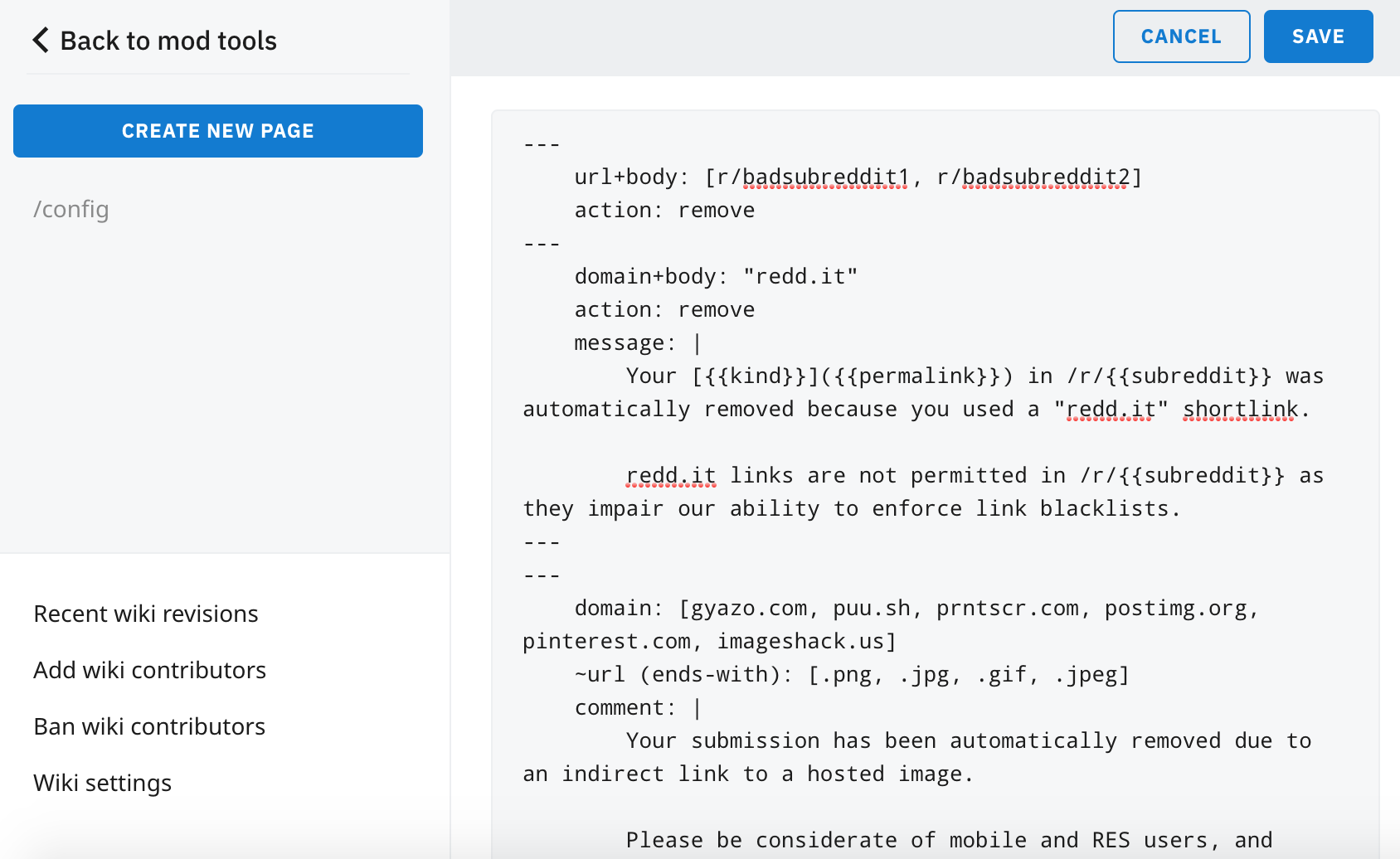}
  \caption{}
  \label{fig:reddit}
\end{subfigure}%
\begin{subfigure}{.5\textwidth}
  \centering
  \includegraphics[width=.8\linewidth]{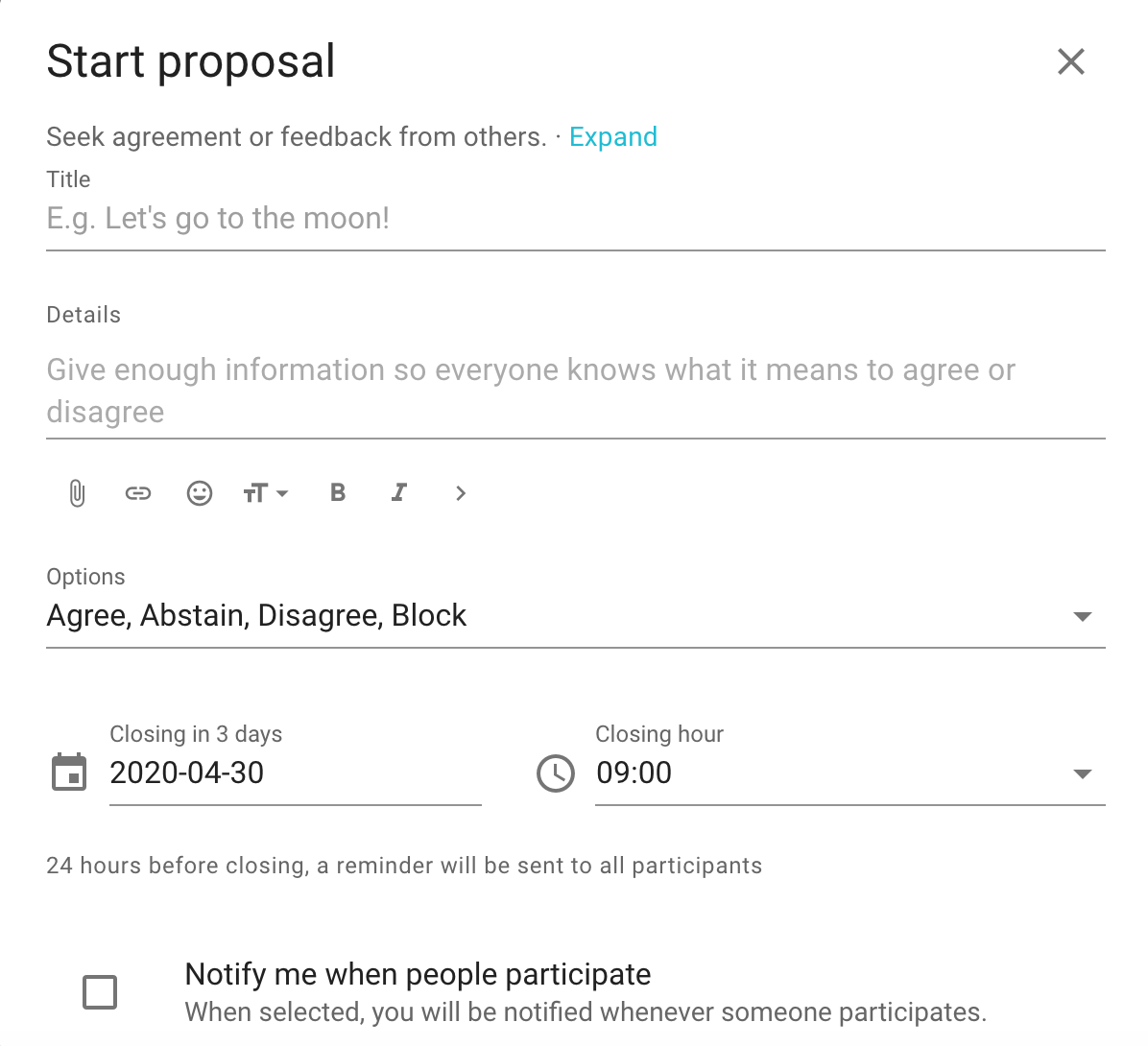}
  \caption{}
  \label{figloomio}
\end{subfigure}
\begin{subfigure}{.5\textwidth}
  \centering
  \includegraphics[width=.8\linewidth]{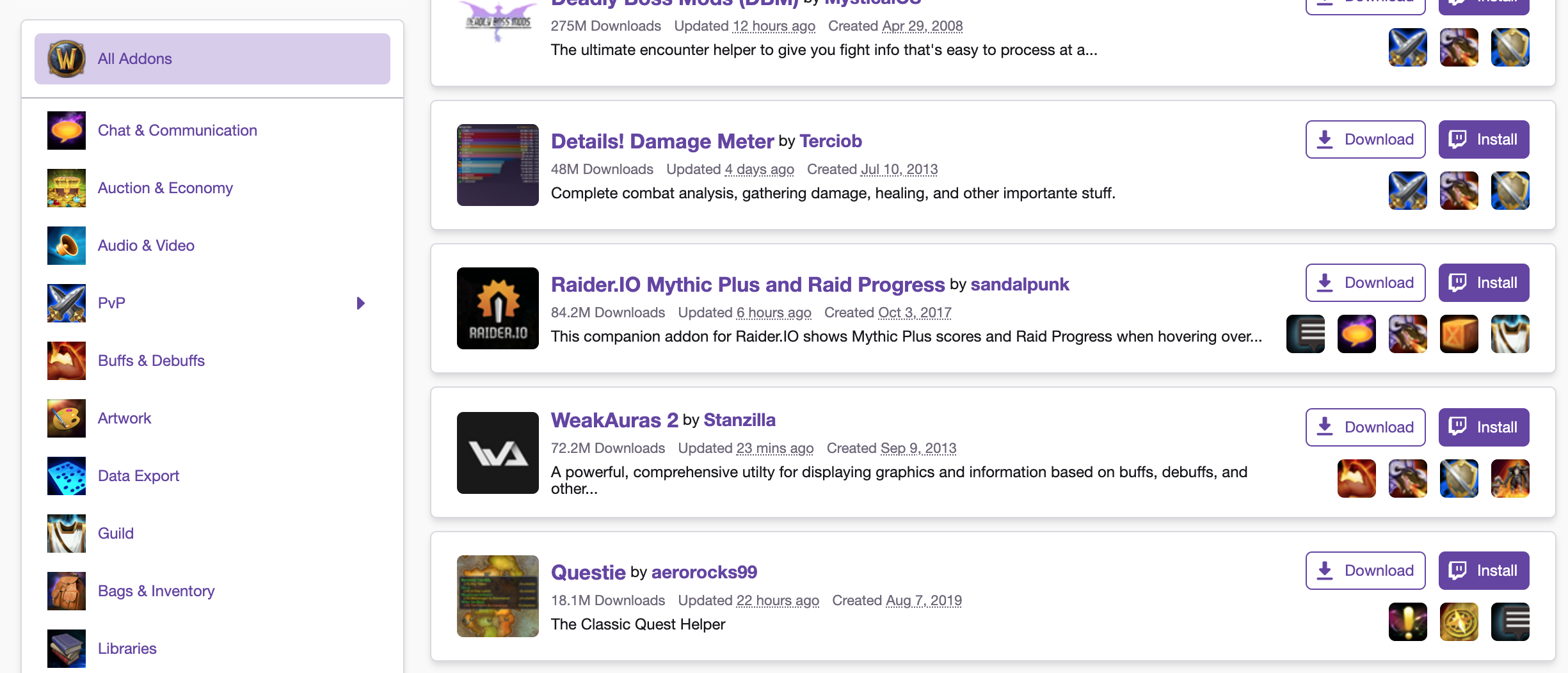}
  \caption{}
  \label{fig:curseforge}
\end{subfigure}%
\begin{subfigure}{.5\textwidth}
  \centering
  \includegraphics[width=.8\linewidth]{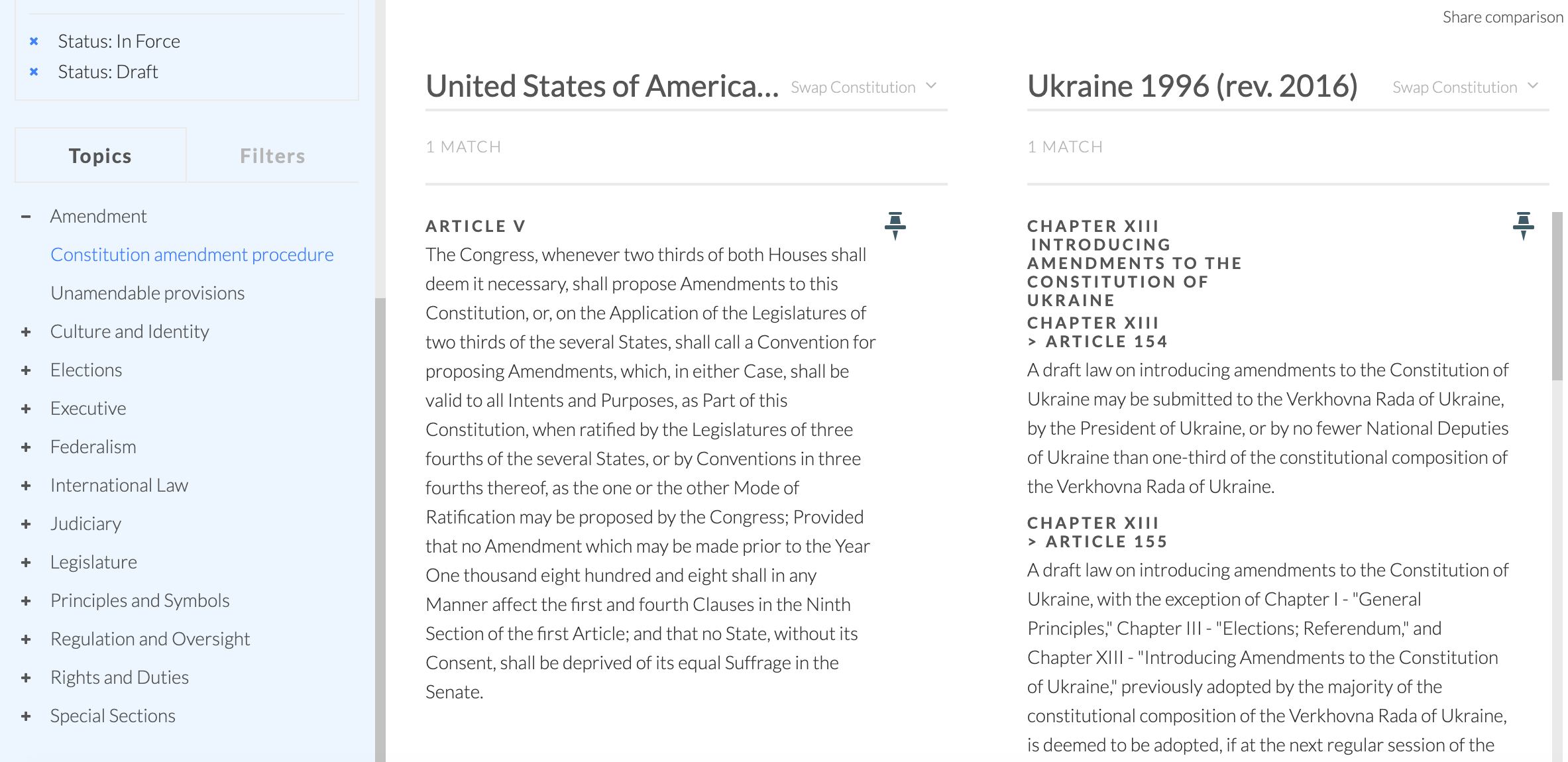}
  \caption{}
  \label{fig:constitute}
\end{subfigure}
\caption{Examples of interfaces analogous to the kinds of configurations possible in Modular Politics. (a) MultiMC, a user-developed, open-source server manager and mod manager for Minecraft, (b) interface for setting rules for Facebook Groups; the rules here are templates suggested by Facebook, (c) Reddit Automoderator, a simple interface for implementing automated content moderation rules, (d) proposal settings on Loomio, a collective decision-making platform, (e) Curseforge, an addon manager and website for addons to several games, including World of Warcraft, (f) the Constitute Project, a collection of historical and in-force constitutions organized by an extensive ontology of terms.}
\label{fig:interfaces}
\end{figure}

Moreover, to enable advanced configuration, Modular Politics would benefit from the use of software development kits (SDKs), which can be both internal or external to a given Platform (Figure \ref{fig:config}). Guilded Age, for example, could implement an internal configuration tool available to server administrators and players. Through an SDK with an intuitive interface, Platform operators and participants alike might browse existing Modules or create their own; they might also adapt, customize, and debug these Modules before importing them into an Instance. SDKs could offer such features as a visual scripting editor~\cite{burnett-visual1999} or a computational language~\cite{wolfram-what2019} to support the development of fine-grained governance structures regardless of a person’s technical skill level. Ensuring that Modules are portable, enabling users to adopt ones created for other platforms, will mean that people lacking the technical skills to create their own Modules can nevertheless choose and adopt tools created by others.

\subsection{Interface and Experience}

Any Instance of Modular Politics must operate within a Platform, which can range in complexity from a simple Web server to a cloud-based multiplayer game or a blockchain protocol. Platform operators will have to implement strategies for defining and circumscribing the role of Modular Politics in the context of the host system, restricting the scope to particular participants and spheres of influence. Operators will also need to define how any particular Instance of Modular Politics interacts with their own systems. For Modular Politics to be effective, it should blend into its environment. It should also support and encourage best practices for accessibility and universal design wherever possible.

Modular Politics does not specify the details of interface or appearance, enabling developers to adapt the system to their needs. The developers of Guilded Age would have wide latitude in specifying how players experience the game, including its governance features (Figure \ref{fig:interfaces}). Yet, for the sake of clarity and consistency, Modular Politics should enforce some basic, shared logics of interaction. The design goal of portability should extend not just to Modules, that is, but also to user experience. Developers should expect to accept certain limits in order to benefit from the cross-platform Module ecosystem of Modular Politics. A person who gains proficiency using a Modular Politics system on one Platform will hence be able to transfer that proficiency to another Platform.

\subsection{Implementation Strategies}

We have outlined in general terms how Modular Politics might operate. But we have not specified the particular form it should take in software. Any implementation should be capable of meeting the goals stated at the outset: modularity, expressiveness, cross-platform portability, and cross-platform interoperability.

One approach to achieving these goals would be to offer Modular Politics solely through one central cloud-based service, which various Platforms would call upon over an API to handle their own governance processes. Such a centralized approach could provide a single, convenient repository of vetted Modules and assure the consistency and reliability of all hosted Instances. Centralization could also help fund the development of Modular Politics, since access to the API might require paying fees to the organization that manages it. Yet some organizations might be reluctant to trust an external authority to provide a service as crucial as governance.

An alternative would be to define Modular Politics itself as a decentralized application running on a distributed-ledger technology (DLT); various Platforms could act as nodes on the protocol without delegating control over these processes to any central authority. But relying on DLTs might also introduce scalability and usability limitations; many such systems have not yet proven ready for widespread adoption. Furthermore, DLTs usually depend on token economics that could impose financial barriers on the adoption of Modular Politics. The economic imperatives of DLTs would also significantly constrain the expressiveness of Modular Politics for diverse types of institutions. While Modular Politics should be implementable on DLTs, it should not presume DLT infrastructure or that of any other specific type of software platform.

We believe the most flexible strategy would be to define an open standard, which can be implemented through a variety of open-source software libraries (perhaps resulting in a full-fledged software framework). The open standard would define the features and behaviors of any Modular Politics system, allowing for interoperability even when the actual implementations differ. Modular Politics could thus enable many kinds of Platforms to develop in tandem, through both replication and direct interaction~\cite{masnick}. The modularity of Modular Politics thereby occurs not just within a particular community but potentially through polycentric relationships among them.

By creating an open standard for a governance protocol, there would be no need to depend on a single organization’s API or rely on the kinds of ``cryptoeconomics'' that DLT systems require---although both would be possible to implement. Guilded Age might depend on both a central server for coordinating gameplay alongside peer-to-peer interactions among servers. Additionally, because our design goals of cross-platform portability and interoperability are more complex and less essential for some use-cases, the standard should allow for implementations that don’t include all the features described here. As with open-source projects like Debian and Python, the development of the standard and its libraries should be community-governed, presumably using Modular Politics itself. Through such distributed collaboration, it is our hope that Modular Politics could become a widespread and extensible layer for governance on the Internet.

\section{Use Cases}   

In order to further elucidate how Modular Politics might operate in practice, we illustrate its scope with two more specific scenarios in which such a system might be employed. The purpose of these hypothetical examples is to illustrate the range of institutions that a successful Modular Politics could serve.

The participants in these scenarios experience action arenas based on a set of pliable, co-determined governance conditions. Participants encounter an interface that enables them to navigate among the various features of the system, take action, and modify the system. Rather than adopting \textit{ad hoc} governance practices within a rigid set of constraints, participants have the capacity to develop and employ tools appropriate to their communities’ needs. Particular Modular Politics systems also have the option of interoperability with other systems, sharing governance mechanisms and corresponding data.

\subsection{Example 1: Social-media moderation}

\begin{figure}
  \centering
  \includegraphics[width=.7\columnwidth]{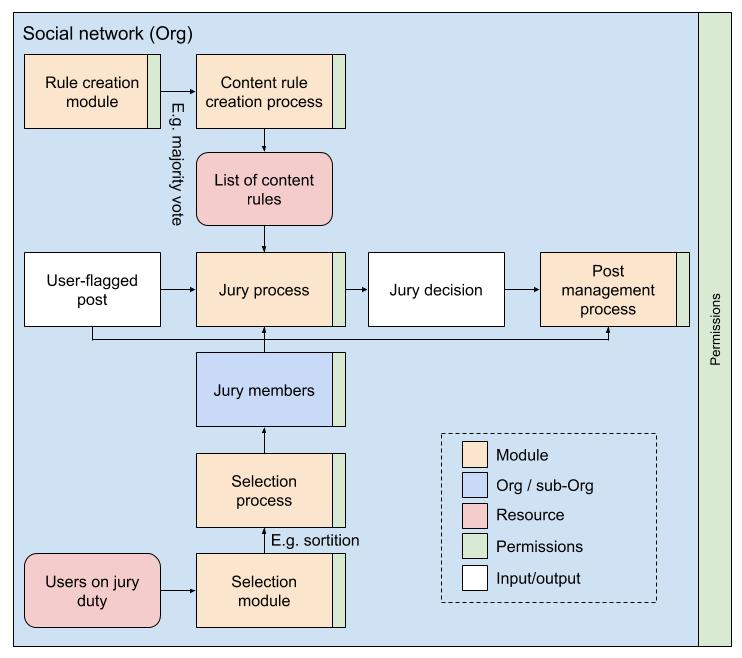}
  \caption{A schematic implementation of the system described in Example 1. Here, a community on social media implements a digital jury system for content moderation.}~\label{fig:jury}
\end{figure}

Consider an affinity-based group on a social-media platform in which several thousand members share and discuss news about sculpture (Figure \ref{fig:jury}). The founders of the group adopted a simple system that enables members of the group to create governance proposals and pass them with a referendum-style vote by a majority of active members.

A controversy arises because some members begin posting pictures of sculptures that other members consider obscene and do not want to see in their feeds. Members begin proposing rules about what kinds of content should not be allowed, and several proposals pass by a majority vote. The founders, who have moderation authority by default, attempt to implement the rules by removing offending posts. But members begin to complain that the founders are being sloppy and non-transparent in how they interpret the rules. One member proposes to add a new tool that implements juries in the group, and the proposal passes. After that, whenever a user flags a post as offensive, it disappears and the platform queries five randomly selected members who previously volunteered for jury duty. If jury members support the removal of a post, they have to specify which rule or rules they believe it has broken. If no members object, the post is removed permanently. Otherwise, it reappears in the group.

This system satisfies most members, but a few raise concerns that the moderation has turned into overly zealous censorship. Although they cannot convince a majority to rescind the system, they do succeed in passing a proposal for a bot that automatically displays statistics on how many posts were flagged and how the juries voted on them. This feedback helps reveal which rules seem most ambiguous to juries, spurring the development of more refined rule definitions and, eventually, lower rates of removal.

\subsection{Example 2: Open-source software project}

\begin{figure}
  \centering
  \includegraphics[width=.8\columnwidth]{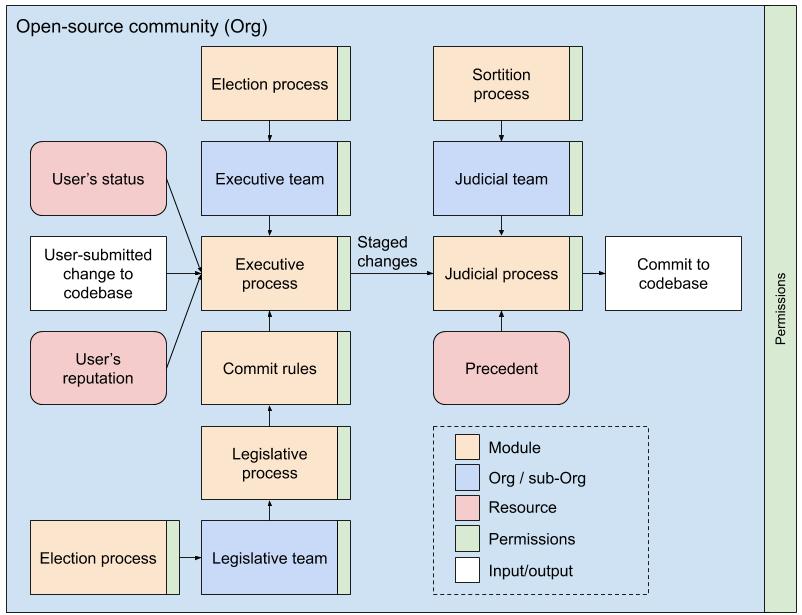}
  \caption{A schematic implementation of the system described in Example 2. Here, an open-source community implements a governance process for code commits involving several modules, ranging from those that specifically govern commits (commit rules) to those that govern who decides the commit rules (election process, sortition process, judicial process).}~\label{fig:oss}
\end{figure}

Consider an open-source software project that produces a popular, privacy-focused mobile operating system (Figure \ref{fig:oss}). Its founder is a developer who comes to be regarded as a ``benevolent dictator for life,'' or BDFL. Although she is generally respected, her unvarnished leadership style frequently distracts the community from its core development work. The project hosts its codebase at BitGit, a decentralized, blockchain-based repository protocol that supports Modular Politics. In one of her moments of exasperation, the BDFL assembles a governance system that resembles a liberal democratic government, with distinct, elected bodies for establishing policies (legislative), implementing those policies (executive), and overseeing the implementation (judicial). To vote or hold office, a developer needs to hold a staking token obtained from making at least one contribution incorporated into the codebase during the past year.

The former BDFL at first refrains from seeking office in the new system, but she doesn’t exactly relinquish leadership. Her charismatic authority continues to hold sway, and the new elected leaders complain that they don’t have sufficient mandate to carry out their roles and expect community support. They introduce a monitoring system into the project’s BitGit interface that shows statistics on participation in the governance process, both for the project as a whole and for any given individual developer. The system also compares these statistics with those of other software projects also using Modular Politics on the BitGit ledger. This puts public pressure on developers to participate more actively in governance, and when they do, they tend to listen more to each other than to the BDFL’s opinions. Eventually, the BDFL runs for a seat on the legislative body and is elected, thus channeling her role through the system she created.

\section{Discussion}
Modular Politics, as it has been presented in this paper, is a strategy for enabling online communities to experiment with a wider variety of governance structures than is typically possible on existing Internet platforms. It draws on the logic of complex, polycentric action arenas as described in IAD research. Beyond simply replicating governance practices found elsewhere, online or offline, the model enables users to investigate novel opportunities for governance in the digital realm. Modular Politics could enact governance models that do not (and perhaps could not) exist in the offline world, inviting platform operators and their users to devise novel organizational practices for themselves.

On the one hand, we hope Modular Politics might enable online communities to adopt democratic practices more often than they presently do. But regardless of the systems they adopt, we suspect there will be benefits to making governance a more explicit feature of communities' experience, so as to prevent them from falling into a ``tyranny of structurelessness''~\cite{freeman-tyranny1972}---a tyranny that has made no exception for technologists~\cite{cohen-1970s2018}. Our aim is not to make a moral or normative case for how online governance should be, but rather to empower all types of communities to develop experiments in governance and practice institutional artisanship.

For researchers, the computational nature of Modular Politics provides a means for studying the emergence and effects of various governance systems, including novel ones. User data has the potential to inform an ongoing conversation with existing social and political theories. The model already provides for reflecting back to users significant data on governance processes. Where appropriate, and according to policies set by platforms and communities themselves, this data could enable social scientists to generate new and more useful taxonomies of governance practices that, if properly analyzed, could contribute to improving the governance practices in various contexts. Beyond the immediate, practical benefits of such research, evidence from Modular Politics could contribute to the development of more robust theories of computational governance.

If implemented as an open standard, rather than as a centralized platform, Modular Politics could contribute to furthering the vision of what has been called ``Web 3.0''---an open and interoperable network of online services that resist the tendencies of centralization that have become so persistent on the Internet today. In the context of governance, this means that a particular community could rely on a variety of integrated services in order to achieve a particular governance structure that spans across multiple platforms. For instance, a community developing open-source software could choose to combine the reputation system of the social news platform Slashdot and moderation standards of a particular group on Reddit in order to assign influence to different users within the community’s governance structure. Deliberation could occur on Discord chat, whereas decisions on the decision-making could be done on the Loomio platform. Once a decision has been made, the implications could be automatically enforced across the relevant platforms. An open standard for governance could thereby contribute to the emergence of specialised services designed to respond to the needs of different communities, while maintaining a high level of integration across these services in order to ensure a cohesive experience of community governance. It is our hope that such a model will inspire users to develop ever more sophisticated governance techniques, which others can adopt in their own contexts.

In addition to fostering new types of online governance innovation and experimentation, we hope to see the effects of Modular Politics extend beyond online communities to political cultures more broadly. If Modular Politics is successful in accelerating the pace of governance innovation through experimentation, the resulting insights can be instantiated in a variety of communities, regardless of the medium of choice. Whereas people typically experience governance in workplaces and governments as fixed and remote, Modular Politics could make the practice of designing and testing governance systems far more widespread, and contribute to the revitalization of civic engagement that the Internet has so long promised and struggled to deliver. While many people's experiences of political participation tend to be limited to such mechanisms as electing representatives or voting in referendums, Modular Politics could expose them to other forms, such as delegative voting systems or sortition. There is surely also a wide spectrum of alternative governance models that have yet to be conceived. Online communities could become powerful laboratories testing new ideas and learning from the failures and successes of other communities, potentially spurring a renaissance in democratic culture.

\section{Limitations} 
Modular Politics, along with our treatment of it here, has a number of limitations.
We begin with limitations of this paper, indicating future work.

First, we have argued for the importance of governance innovation throughout this article but have not proven (conceptually or empirically) that it would be easier to innovate in Modular Politics than otherwise. We believe that modularity and portability, in particular, will accelerate innovation, but there are many factors (e.g., software usability, software variability or malleability, programming language design, simulation and inference tools) that could help or hinder the effectiveness of Modular Politics as a tool for innovation.

Second, we have not defined exactly what kinds of new governance systems and structures would and would not be expressible in Modular Politics. For example, future research that takes a more mathematical approach might define an explicit space of governance structures and prove that Modular Politics is able to recover all elements in that space.

Third, although Modular Politics is intended to support heterogeneous institutional forms, our formulation surely bears built-in political biases that we have yet to interrogate. All technological artifacts such as this carry political qualities, whether because they tend to be used for certain political purposes or because their design inclines toward a certain political system or philosophy \cite{winner2004artifacts}. One-size-fits-all technical solutions run the risk of imposing a false universalism \cite{milan-big2019}. We hope to learn from early user testing how to ensure that Modular Politics facilitates bottom-up governance innovation rather than imposing norms. We also have yet to explore how commercialization and economic incentives might affect the use of Modular Politics in practice.

There are also limitations of Modular Politics as a model, indicating fundamental constraints and issues of scope.

First, Modular Politics is not intended as a proposal for offline community governance. While we believe that online experiments will eventually teach us much about about offline governance, we have chosen to narrow the scope of our present inquiry to communities that exist and interact mostly online through digital platforms and messaging services. 

Second, Modular Politics does not constitute or aspire to be a complete theory of governance. It is, rather, a proposal for tools that might support various forms of governance within a digital ``state of nature,'' however fictional such a state inevitably is. The model can, however, serve as the basis of theoretical claims and empirical tests about the system and the behavior of its users. We also understand our proposal as a complement to various efforts that introduce more accountable governance into the ownership structures, business models, and public policies for online platforms~\cite{scholz-ours2016}.

Finally, and of particular importance, we emphasize once again that constructing an effective system of governance requires more than just software. Cultural and contextual factors are also essential~\cite{bollier-patterns2015}. Modular Politics does not specify, but depends upon, a broader "civic sphere," by which we mean the confluence of tools, culture, and context through which participants interact. A healthy civic sphere is one in which participants feel confidence that the governance process is meaningfully accountable to them~\cite{decaro-humanistic2018}, with some kind of shared norms, appropriate information flows, and the power to influence decisions. Crafting the civic sphere is the job of platform operators, community leaders, and community members as a whole. For example, a particular community might adopt language that makes governance activities appear fun and silly, or alternatively grave and serious. Policies might require prospective participants to agree to act in good faith, or other prerequisites, before joining the community. Platform operators might choose to implement Modular Politics in ways that bear biases toward more or less democratic practices. Users in positions of authority might choose to emphasize negative sanctions for rule-breakers or positive reinforcement. These kinds of choices are vitally important, and Modular Politics does not prescribe them. Modular Politics, as with any procedural or computational system, is no replacement for other aspects of a healthy civic sphere.

\section{Conclusion}

The tools available for governance in online communities are currently limited, inflexible, and ill-equipped for governance innovation. Modular Politics seeks to encourage such innovation through a dynamic, flexible model for online governance, through which community members can engage in creating and experimenting with a variety of different governance techniques. As such, Modular Politics could accelerate and proliferate innovation in governance design well beyond what occurs in offline systems, which carry considerable burdens of inertia and path dependence. Ultimately, a successful open standard for governance could contribute to making creative and responsive governance a more widespread norm—both online and offline.

This paper is a preliminary step toward that goal. Much research and experimentation remains in order to define a standard that is both durable and attractive for platforms to adopt, which would require specifying Modular Politics in much greater detail than we have done here. We hope at least to have spurred interest in the challenges to come.

\begin{acks}
The authors are grateful for substantive feedback from Kris Jones, Daniel Kronovet, nie ls, Nick Meyne, Aviv Ovadya, David Rozas Domingo, Abbey Titcomb, and Glen Weyl. Nathan Schneider acknowledges support from a fellowship with the Open Society Foundations.
\end{acks}

\bibliographystyle{ACM-Reference-Format}
\bibliography{base}

\appendix

\end{document}